\documentclass{article}
\usepackage{spconf,amsmath,graphicx}
\usepackage{comment,multirow,multicol,balance,pifont,siunitx,circledsteps,scalerel,moresize,tipa,color,amsfonts,amssymb,setspace,bm,amsmath}

\title{DISCONTSE: SINGLE-STEP DIFFUSION SPEECH ENHANCEMENT \\BASED ON JOINT DISCRETE AND CONTINUOUS EMBEDDINGS}
\name{Yihui Fu,
      Tim Fingscheidt}
\address{\hspace{-3mm}Institute for Communications Technology, Technische Universität Braunschweig, Braunschweig, Germany}
\begin{document}
\ninept
\maketitle
\begin{abstract}
Diffusion speech enhancement on discrete audio codec features gain immense attention due to their improved speech component reconstruction capability. However, they usually suffer from high inference computational complexity due to multiple reverse process iterations. Furthermore, they generally achieve promising results on non-intrusive metrics but show poor performance on intrusive metrics, as they may struggle in reconstructing the correct phones. In this paper, we propose \texttt{DisContSE}, an efficient diffusion-based speech enhancement model on joint discrete codec tokens and continuous embeddings. Our contributions are three-fold. First, we formulate both a discrete and a continuous enhancement module operating on discrete audio codec tokens and continuous embeddings, respectively, to achieve improved fidelity and intelligibility simultaneously. Second, a semantic enhancement module is further adopted to achieve optimal phonetic accuracy. Third, we achieve a \textit{single-step efficient reverse process} in inference with a novel \textit{quantization error mask initialization} strategy, which, according to our knowledge, is the first successful single-step diffusion speech enhancement based on an audio codec. Trained and evaluated on URGENT 2024 Speech Enhancement Challenge data splits, the proposed \texttt{DisContSE} excels top-reported time- and frequency-domain diffusion baseline methods in PESQ, POLQA, UTMOS, and in a subjective ITU-T P.808 listening test, clearly achieving an overall top rank.
\end{abstract}

\vspace{-1mm}
\section{Introduction}
\vspace{-1mm}
Diffusion-based speech enhancement (SE) has gained widespread attention. Complex-valued frequency-domain representation is commonly adopted for score-based diffusion models due to the better phase reconstruction ability and intelligibility \cite{ richter2023speech, lay2023reducing,richter2024investigating}. However, multiple iterations in the reverse process greatly increase the computational complexity during inference. Recently, SE on \textit{discrete} audio codec features with masked token estimation method has received increasing attention because of its outstanding denoising ability and fidelity \cite{wang2024selm,kang2025llaseg,MaskSR,yang2024genhancer,zhang2025anyenhance}. On the other hand, SE on \textit{continuous} embeddings is proven to achieve good phonetic accuracy and speaker preservation ability \cite{liu2024audiosr,guimaraes2025ditse}. However, all these methods still suffer from high inference complexity due to multiple reverse iterations. Furthermore, the underlying relationship between the discrete and continuous embeddings is still to be investigated. In this paper, we propose $\texttt{DisContSE}$, an \textit{efficient} diffusion-based SE model on \textit{joint discrete and continuous} embeddings. We first integrate a discrete enhancement module and a supervised continuous enhancement module to deliver enhancement on discrete audio codec tokens and continuous embeddings, respectively, with the features extracted by the pre-trained descript audio codec (\texttt{DAC}) \cite{kumar2023high}. Then, a supervised semantic enhancement module based on \texttt{WavLM}-encoded features \cite{chen2022wavlm} is further adopted to increase the phonetic accuracy. Finally, a novel \textit{single-step efficient reverse process} is presented, integrating the continuous enhancement module output and a newly proposed \textit{quantization error mask initialization} strategy.

\section{Methods}

This section describes training and inference of our proposed \texttt{DisConSE}.

\begin{figure*}[ht!]
    \centering
    \raisebox{0cm}{\hspace{-0.38cm} 
    \includegraphics[width=2.05\columnwidth]{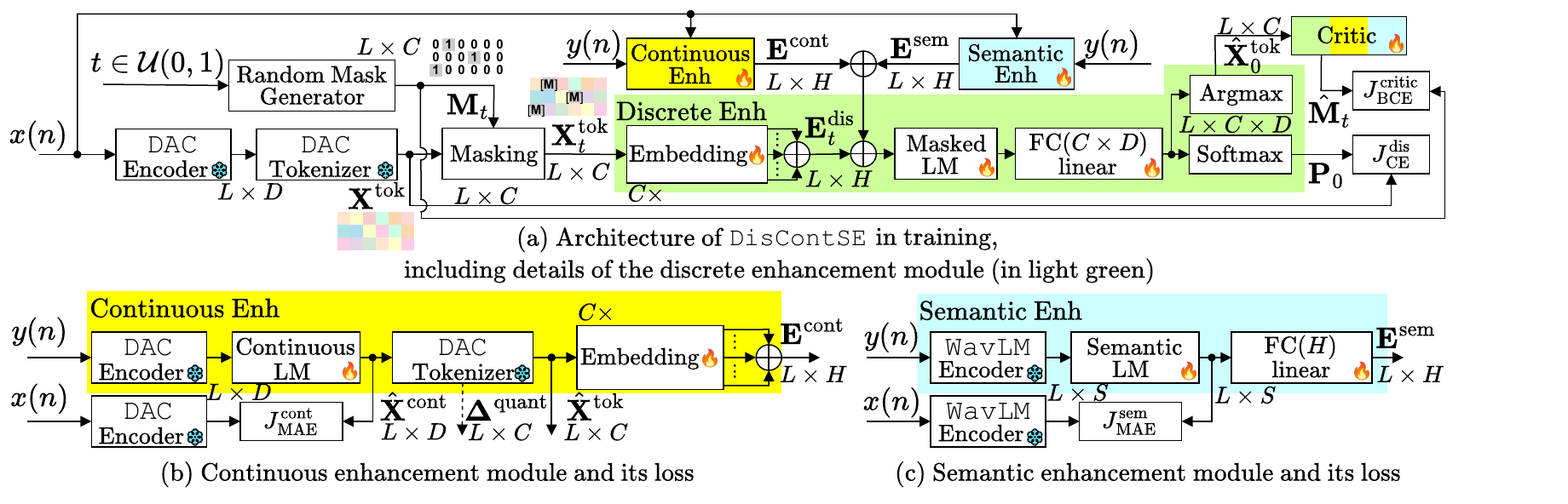}}
    \caption{(a) Architecture of the \textbf{proposed \texttt{DisContSE}} and its \textbf{training} strategy. 
    (b) Block diagram of the continuous enhancement module and its MAE loss. 
    (c) Block diagram of the semantic enhancement module and its MAE loss. 
    }
    \label{fig:discoutse_train}
\end{figure*}

\begin{figure}[ht!]
    \centering
    \raisebox{0cm}{\hspace{-0.15cm} 
    \includegraphics[width=1.0\columnwidth]{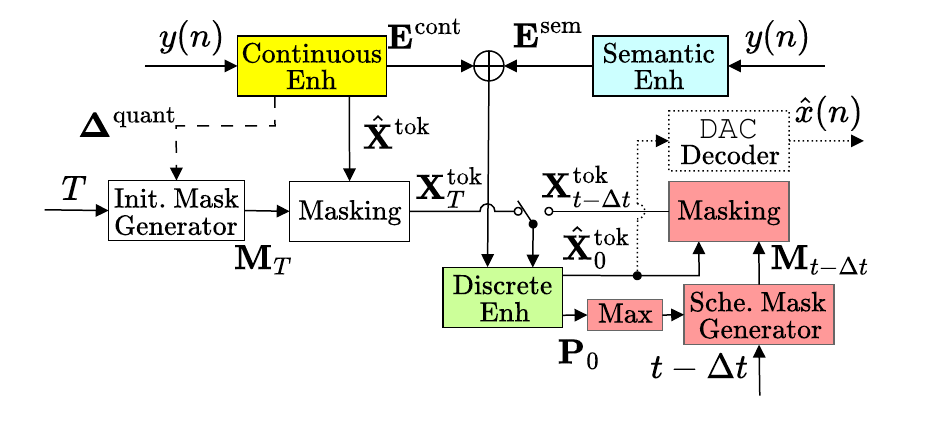}}
    \caption{\textbf{Inference} of the \textbf{proposed \texttt{DisContSE}}.}
    \label{fig:discoutse_infer}
\end{figure}

\textbf{Training}: As shown in Fig. \ref{fig:discoutse_train} (a), the \texttt{DAC} encoder and \texttt{DAC} tokenizer \cite{kumar2023high} are adopted to extract the discrete tokens $\mathbf{X}^{\textrm{tok}}=(X^{\textrm{tok}}_{\ell,c}) \in \{0,1,...,D\!-\!1\}^{L\times C}$ from the clean speech $x(n)$, where $L$ represents the token sequence length, $C$ the number of codebook entries, and $D$ the codebook size. \texttt{MaskGIT} \cite{chang2022maskgit} is adopted as a random mask generator, which takes $t\in \mathcal{U} (0,1)$ as the input to generate the mask $\mathbf{M}_t=(M_{t, \ell,c}) \in \{0,1\}^{L\times C}$, within which an amount of $\left \lfloor \textrm{sin}(\frac{\pi t}{2})\cdot L\cdot C \right \rfloor$ randomly chosen elements fulfill $M_{t,\ell,c}=1$, all others are $M_{t,\ell,c}=0$. The masked tokens $\mathbf{X}^{\textrm{tok}}_t=(X_{t,\ell,c}^{\textrm{tok}})\in \{0,1,...,D\}^{L\times C}$ are obtained by 
\begin{equation}
\label{masking}
X_{t,\ell,c}^{\textrm{tok}}=[\mathrm{M}] \cdot M_{t,\ell,c}  + X_{\ell,c}^{\textrm{tok}}\cdot (1-M_{t,\ell,c}),
\end{equation}
with $[\mathrm{M}]$ being a special fixed token with value $D$.
The discrete enhancement module contains $C$ different parallel embedding layers with embedding dimension $H$ applied to each codebook entry with index $c$ and token sequence with index $\ell$, with summation output $\mathbf{E}^{\textrm{dis}}_t \in \mathbb{R}^{L\times H}$. The embeddings $\mathbf{E}^{\textrm{cont}}$ and $\mathbf{E}^{\textrm{sem}}$, estimated by the continuous enhancement and semantic enhancement modules, respectively, are added to $\mathbf{E}^{\textrm{dis}}_t$ to deliver the input to the masked language model (LM) \cite{chang2022maskgit,MaskSR,yang2024genhancer,zhang2025anyenhance}. The masked LM consists of cascaded transformer blocks with dimension $H$. After a fully connected (FC) linear layer $\textrm{FC}(C\times D)$ applied to each token with index $\ell$ in the sequence, a cross-entropy loss 
\begin{equation}
\label{dis ce unmasked}_{}
J_{\textrm{CE}}^{\textrm{dis}}(\mathbf{P}_{0},\textbf{X}^{\textrm{tok}}) =
 J_{\textrm{CE}}(\mathbf{P}_{0},\textbf{X}^{\textrm{tok}})|_{M_{t,\ell,c}=1}
\end{equation}
is applied between softmax output $\mathbf{P}_0=(P_{0,\ell,c,d})\in \mathbb{R}^{L\times C \times D} $ and clean speech tokens $\mathbf{X}^{\textrm{tok}}$, where the gradients are calculated only on the masked positions ($M_{t,\ell,c}=1$) \cite{MaskSR,yang2024genhancer,zhang2025anyenhance}.
On the masked positions, the estimated tokens $\hat{\mathbf{X}}_0^{\textrm{tok}}$ are derived from the argmax operation over the $D$-dimension of the $\textrm{FC}(C\times D)$ output, while the tokens on the unmasked positions are copied from $\mathbf{X}^{\textrm{tok}}$. According to \cite{zhang2025anyenhance}, a self-critic sampling strategy is further adopted, which reuses the \texttt{DisContSE} as a discriminator to evaluate whether a token is real or generated by the model. By substituting the $\textrm{FC}(C\times D)$ linear layer by an $\textrm{FC}(C)$ linear layer and substituting softmax by sigmoid in the discrete enhancement, and reusing all other parameters, the estimated mask $\hat{\mathbf{M}}_t$ is optimized by binary cross-entropy loss $J_\textrm{BCE}^{\textrm{critic}}(\hat{\mathbf{M}}_t, \mathbf{M}_t)$.
During training, the parameters of the \texttt{DAC} encoder, \texttt{DAC} tokenizers, and \texttt{WavLM} encoder are kept frozen. We employ the overall loss
\begin{equation}
\label{loss sum}
J= J_\textrm{CE}^{\textrm{dis}} + J_\textrm{BCE}^{\textrm{critic}}+ J_{\textrm{MAE}}^{\textrm{cont}} + J_{\textrm{MAE}}^{\textrm{sem}}.
\end{equation}

Figs. \ref{fig:discoutse_train} (b) and (c) represent the continuous enhancement module and semantic enhancement module, respectively. 
The continuous enhancement module, acting as a discriminative enhancement module, delivers continuous embeddings enhancement on the noisy speech $y(n)$ being \texttt{DAC}-encoded, using a continuous LM \cite{li2025speech}, which consists of two FC linear layers $\textrm{FC}(H)$ and $\textrm{FC}(D)$, and cascaded transformer blocks with dimension $H$ lying between. Beyond \cite{yang2024genhancer}, we employ an MAE loss $J_{\textrm{MAE}}^{\textrm{cont}}$ to optimize the enhanced continuous embeddings $\hat{\mathbf{X}}^{\textrm{cont}}\in \mathbb{R}^{L\times D}$. Thus, the optimized $\hat{\mathbf{X}}^{\textrm{cont}}$, as a pre-enhanced speech feature, provides a reliable initial state for achieving efficient single-step diffusion. Discrete enhanced tokens $\hat{\mathbf{X}}^{\textrm{tok}}$ are derived from $\hat{\mathbf{X}}^{\textrm{cont}}$ by \texttt{DAC} tokenizers. Then $\mathbf{E}^{\textrm{cont}}\in \mathbb{R}^{L\times H}$, a newly proposed conditional information, is estimated by summing the outputs of $C$ different parallel embedding layers. Note that for parameter efficiency, the embedding layers in the discrete enhancement module share weights pairwise with those in the continuous enhancement module.

The semantic enhancement module, inspired by \cite{MaskSR,zhang2025anyenhance}, delivers semantic feature enhancement on the noisy speech $y(n)$ being \texttt{WavLM}-encoded \cite{chen2022wavlm} using a semantic LM, which consists of two FC linear layers $\textrm{FC}(H)$ and $\textrm{FC}(S)$, and cascaded transformer blocks with dimension $H$ lying between. Beyond \cite{yang2024genhancer}, the MAE loss $J_{\textrm{MAE}}^{\textrm{sem}}$ is adopted to optimize the enhanced semantic features. An $\textrm{FC}(H)$ linear layer is further adopted to estimate $\mathbf{E}^{\textrm{sem}}\in \mathbb{R}^{L\times H}$.

\textbf{Inference}: As shown in Fig. \ref{fig:discoutse_infer}, during inference of the proposed $\texttt{DisContSE}$, initial mask $\mathbf{M}_{T}=(M_{T, \ell,c})\in\{0,1\}^{L\times C}$ is generated according to the initial diffusion time index $T\leqslant 1.0$, where $T=1.0$ represents fully masked initialization \cite{MaskSR,yang2024genhancer,zhang2025anyenhance}, i.e., $M_{T, \ell,c}=1$. Furthermore, we propose two initial mask generating strategies with $T<1.0$:  

(i) \textit{Quantization error mask initialization}, which takes the \texttt{DAC} tokenizer (dashed arrow in Figs.\ \ref{fig:discoutse_train} (b) and \ref{fig:discoutse_infer}, left) quantization error matrix $\boldsymbol\Delta ^{\textrm{quant}} \in\mathbb{R}_{+}^{L\times C}$, with the MSE quantization error of each of the $C$ quantizers for each token sequence index $\ell$ as elements. Then, in the positions of the largest $\left \lfloor \textrm{sin}(\frac{\pi T}{2})\cdot L\cdot C \right \rfloor$ entries in $\boldsymbol\Delta ^{\textrm{quant}}$, we set mask $\mathbf{M}_T$ elements to $M_{T, \ell,c}=1$.

(ii) \textit{Random mask initialization} (no dashed arrows), which randomly assigns $M_{T, \ell,c}=1$ on $\left \lfloor \textrm{sin}(\frac{\pi T}{2})\cdot L\cdot C \right \rfloor$ positions in $\mathbf{M}_T$, equaling the mask generation process in training.

By applying $\mathbf{M}_{T}$ to the discrete enhanced tokens $\hat{\mathbf{X}}^{\textrm{tok}}$, in analogy to (\ref{masking}), the initial state $\mathbf{X}^{\textrm{tok}}_T$ is calculated and fed into the discrete enhancement module (switch in left position), together with $\mathbf{E}^{\textrm{cont}}\!+\mathbf{E}^{\textrm{sem}}$. In each iteration of the reverse process with index $t$, $\mathbf{P}_0$ and $\hat{\mathbf{X}}^{\textrm{tok}}_0$ are estimated by the discrete enhancement module. 

If multiple reverse iterations are executed (light red boxes), a max operation is applied to $\mathbf{P}_0$ on its $D$-dimension. Then, the smallest $\left \lfloor \textrm{sin}(\frac{\pi (t-\Delta t)}{2})\cdot L\cdot C \right \rfloor$ entry positions, which are considered as estimated low-confidence tokens \cite{chang2022maskgit}, are adopted to remask $\hat{\mathbf{X}}^{\textrm{tok}}_0$, using $\mathbf{M}_{t-\Delta t}$ estimated by the scheduled mask generator, to generate $\mathbf{X}^{\textrm{tok}}_{t-\Delta t}$ as the next iteration's input to the discrete enhancement module (switch in right position). Here $\Delta t=T/N$, with $N$ being the number of the reverse steps during inference. 

In the last iteration (or the only one in our proposed single-step reverse process), the discrete enhancement module estimates $\hat{\mathbf{X}}^{\textrm{tok}}_0$, which is further input to the \texttt{DAC} decoder to estimate the enhanced speech $\hat{x}(n)$, shown by dotted arrows on the right side in Fig.\ \ref{fig:discoutse_infer}.

\begin{table*}[!t]
\centering
\caption{Performance of the proposed \texttt{DisContSE} method ($T=0.1$) and some popular generative SE models on $\mathcal{D}^{\textrm{test}}$. P.808 MOS is obtained on a 50-sample subset of $\mathcal{D}^{\textrm{test}}$.  Best performance in the last two table segments on each metric is in bold, second best is underlined.}
\scalebox{0.94}
{\hspace{-0.18cm}
{
\begin{tabular}{l|c|c@{\hskip 6pt}c@{\hskip 6pt}c@{\hskip 6pt}c@{\hskip 6pt}c@{\hskip 6pt}c@{\hskip 6pt}c@{\hskip 6pt}c@{\hskip 6pt}c@{\hskip 6pt}c@{\hskip 6pt}c@{\hskip 6pt}c}
\hline
Method  & \multicolumn{1}{c|}{Type}    & PESQ  & POLQA & \begin{tabular}[c]{@{}c@{}}DNS\\ MOS\end{tabular} & NISQA & UTMOS & ESTOI  & LPS &SBScore &SpkSim & \begin{tabular}[c]{@{}c@{}}WAcc\\ (\%)\end{tabular} & MOS  & \begin{tabular}[c]{@{}c@{}}Overall\\ rank\end{tabular}$\downarrow$ \\ \hline\hline
Noisy &\multicolumn{1}{c|}{-}   &1.88 &2.17 &1.91 &1.66 &1.87 &0.67 &0.72 &0.71 &0.76 &79.91 &2.17 &8.36 \\
Clean &\multicolumn{1}{c|}{-}   &4.50 &4.71 &3.12 &3.87 &3.35 &1.00 &1.00 &1.00 &1.00 &89.08 &3.95 &- \\ 
Clean-DAC &\multicolumn{1}{c|}{-}   &4.31 &4.38 &3.17 &3.98 &3.31 &0.95 &0.95 &0.98 &0.77 &87.74 &3.92 &- \\ \hline\hline
\texttt{SGMSE+} \cite{richter2023speech}  &G30   &2.75 &2.98 &2.79 &3.61 &2.74 &0.78 &0.79 &0.79 &\textbf{0.79} &73.78 &3.49 &6.27 \\
\texttt{BBED} \cite{lay2023reducing}  &G30   &2.54 &2.62 &3.05 &3.94 &3.01 &0.71 &0.69 &0.75 &0.67 &63.71 &3.48 &7.36 \\
\texttt{SB} \cite{richter2024investigating}  &G30  &2.57 &2.86 &\textbf{3.21} &3.61 &3.07 &0.79 &0.80 &\underline{0.82} &0.57 &75.39 &\underline{3.73} &4.82 \\
\texttt{CRP} \cite{lay2024single} &G1  &3.10 &3.01 &3.08 &3.89 &3.04 &\textbf{0.81} &\textbf{0.84} &\underline{0.82} &0.71 &\textbf{78.90} &3.71 &\underline{3.36} \\
\texttt{CDiffuSE} \cite{lu2022conditional} &G1  &2.40 &2.43 &2.66 &2.43 &2.24 &0.58 &0.75 &0.75 &0.53 &\underline{76.45} &2.84 &8.45 \\
\texttt{StoRM} \cite{lemercier2023storm}   &D+G50  &2.94 &3.02 &3.15 &\underline{4.02} &2.95 &0.79 &0.79 &0.80 &\underline{0.77} &72.76 &3.67 &4.82 \\
\texttt{Universe++} \cite{scheibler2024universal}  &D+G8  &3.09 &3.23 &3.14 &\textbf{4.03} &3.04 &\underline{0.80} &0.79  &0.81 &0.60 &73.06 &\underline{3.73} &4.18 \\ \hline\hline
\texttt{DisContSE} (\textbf{prop.}) ($\hat{\mathbf{X}}^{\textrm{tok}}_0$)  & D+G1         &\textbf{3.14} &\textbf{3.25} &\underline{3.19} &3.85 &\textbf{3.13} &\underline{0.80} &\underline{0.82} &\textbf{0.84} &0.60 &75.50 &\textbf{3.75} &\textbf{2.36}  \\ 
~Continuous Enh only ($\hat{\mathbf{X}}^{\textrm{tok}}$) & D          &\underline{3.12} &\underline{3.24} &3.15 &3.76 &\underline{3.10} &\underline{0.80} &\underline{0.82} &\textbf{0.84} &0.59 &74.71 &3.68 &3.55 \\ 
\hline
\end{tabular}
}
}
\label{tab:performance}
\end{table*}

\section{Experimental Setup}
\vspace*{-3mm}
\subsection{Networks}
\vspace*{-1mm}
In the proposed \texttt{DisContSE}, the pretrained 16 kHz \texttt{DAC}\footnote{https://github.com/descriptinc/descript-audio-codec} and \texttt{WavLM}\footnote{https://huggingface.co/docs/transformers/model\_doc/wavlm} weights are adopted. For \texttt{DAC}, both the encoder output dimension and the codebook size are $D\!=\!1024$, and the number of codebooks is $C\!=\!12$. For \texttt{WavLM}, the 6th encoder layer's output is adopted \cite{yao2025gense} with dimension $S\!=\!1024$.
In the discrete enhancement and the continuous enhancement, both the masked LM and the continuous LM consist of 8 transformer blocks, while the semantic LM in the semantic enhancement contains 4 transformer blocks. All transformer blocks have $H\!=\!512$ dimensions and 4 heads. The total number of trainable parameters in our \texttt{DisContSE} is 81.4 M. The number of frozen parameters of \texttt{DAC} and \texttt{WavLM} encoder are 74.2 M and 158.3 M, respectively.

\vspace*{-3mm}
\subsection{Data and Metrics}
\vspace*{-1mm}
We train and evaluate our proposed \texttt{DisContSE} on the large-scale diverse-source URGENT 2024 Speech Enhancement Challenge data splits \cite{zhang2024urgent}, however, excluding the CommonVoice 11.0 English portion \cite{CommonVoice-Ardila2020} due to its occasional background noise \cite{zhang2024urgent}. For training data $\mathcal{D}^{\textrm{train}}$ and validation data $\mathcal{D}^{\textrm{val}}$, we mainly follow the strategies in the challenge\footnote{https://github.com/urgent-challenge/urgent2024\_challenge}, in which 3 kinds of distortions are adopted: additive noise with and without reverberation, and additive noise with clipping. We do not adopt bandwidth limitation distortion from the challenge. The signal-to-noise ratio (SNR) range is [-5, 20] dB. An active speech level with a range [-36, -16] dB is adopted before mixing speech and noise data \cite{ITUSRN}. All waveforms are downsampled to 16 kHz. We configure a 634.5 h training set $\mathcal{D}^{\textrm{train}}$ and a 32.7 h validation set $\mathcal{D}^{\textrm{val}}$, without speech, speakers and noise overlapping. We adopt the officially published non-blind test set of the challenge, excluding the bandwidth limitation waveforms, and downsample it to 16 kHz, leading to our test set $\mathcal{D}^{\textrm{test}}$ with 661 waveforms.

\begin{figure}[h!]
\vspace*{-5mm}
    \centering
    \raisebox{0cm}{\hspace{-0.15cm}
    \includegraphics[width=1.05\columnwidth]{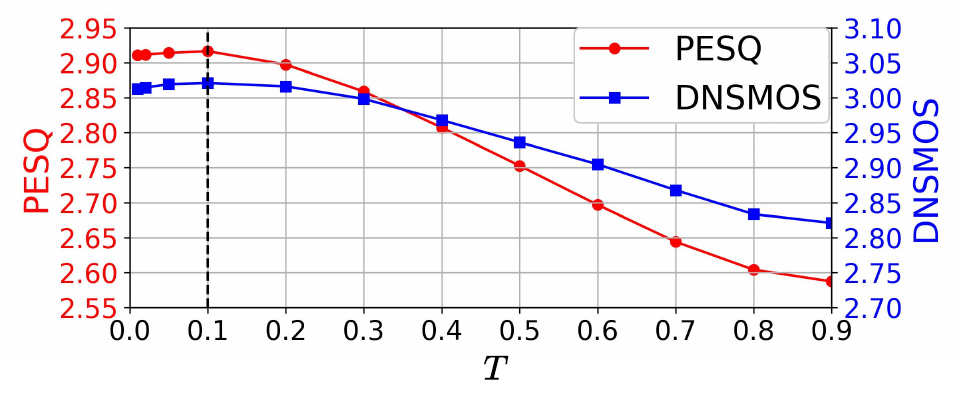}}
    \vspace*{-10mm}
    \caption{\texttt{DisContSE} performance on $T$ in the \textbf{single-step reverse process} (inference) on $\mathcal{D}^{\textrm{val}}$.}
    \label{fig:pesq_dnsmos_T}
\end{figure}
We employ intrusive SE metrics, including PESQ \cite{rix2001perceptual}, POLQA \cite{beerends2013perceptual}, and ESTOI \cite{jensen2016algorithm}, non-intrusive SE metrics, including DNSMOS \cite{reddy2022dnsmos}, NISQA \cite{mittag2021nisqa}, and UTMOS \cite{saeki22c_interspeech}, downstream-task-independent metrics, including Levenshtein phone similarity (LPS ~=~1~-~Levenshtein phone distance (LPD)) \cite{pirklbauer2023evaluation}, as a phone fidelity metric being sensitive to hallucinations, and SpeechBERTScore (SBScore) \cite{SBScore}, and downstream-task-dependent metrics, including SpkSim \cite{zhang2024urgent} and WAcc \cite{zhang2024urgent}. We also perform a subjective listening test to estimate the mean opinion score (MOS) of each model, strictly following ITU-T Rec.\ P.808 \cite{ITU-P808, Open-Naderi2020}. In line with the setup of the URGENT 2024 Challenge \cite{zhang2024urgent}, we select a 50-sample subset of $\mathcal{D}^{\textrm{test}}$, whereby each sample is evaluated by eight different listeners who are self-reported native English speakers. The overall rank is calculated based on the average rank of each metric. 

\vspace{-3 mm}
\subsection{Model Training}
\vspace{-1 mm}
The proposed \texttt{DisContSE} is trained on 4 \texttt{Nvidia H100} with a batch size of 48 for 300 K steps, taking about 3.5 days. The AdamW optimizer is adopted. The learning rate is 0.00025 with 4 K steps of warmup. During training, the parameters of \texttt{DAC} encoder, \texttt{DAC} tokenizer, and \texttt{WavLM} encoder are frozen.
We report some popular generative speech enhancement models as baselines, including \texttt{SGMSE+} \cite{richter2023speech}, \texttt{BBED} \cite{lay2023reducing}, \texttt{SB} \cite{richter2024investigating}, \texttt{CRP} \cite{lay2024single}, \texttt{CDiffuSE} \cite{lu2022conditional}, \texttt{StoRM} \cite{lemercier2023storm}, and \texttt{Universe++} \cite{scheibler2024universal}. For fair comparison, all these models are trained for 1 M steps with batch size 16 on the same dataset $\mathcal{D}^{\textrm{train}}$ as \texttt{DisContSE}. The best setups in the corresponding papers are taken. We perform inference with the number of steps of the best performance in the corresponding papers. For \texttt{CRP}, we use their \texttt{SB} model trained by 950 K steps as the first stage model to deliver the second stage training for 50 K steps, according to the ratio of the 
number of training steps of the two stages in \cite{lay2024single}.

\begin{table*}[t!]
\centering
\caption{Ablation study on the inference strategies ($\Circled{1}$ to $\Circled{5}$) and training strategies ($\Circled{6}$ to $\Circled{9}$), evaluated on $\mathcal{D}^{\textrm{val}}$.}
\scalebox{0.92}
{\hspace{-0.18cm}
{
\begin{tabular}{l|c|c|c@{\hskip 6pt}c@{\hskip 6pt}c@{\hskip 6pt}c@{\hskip 6pt}c@{\hskip 6pt}c@{\hskip 6pt}c@{\hskip 6pt}c@{\hskip 6pt}c@{\hskip 6pt}c}
\hline
Method &$T$   & \multicolumn{1}{c|}{Type}  & PESQ & POLQA & DNSMOS & NISQA & UTMOS & ESTOI  &LPS &SBScore &SpkSim & \begin{tabular}[c]{@{}c@{}}WAcc\\ (\%)\end{tabular} \\ \hline\hline
Noisy &- &\multicolumn{1}{c|}{-} &1.71 &1.87 &1.65 &1.40 &1.65 &0.60 &0.63 &0.63 &0.69 &85.45 \\ \hline\hline
\texttt{DisContSE}  & & & & & & & & & & & &   \\ 
~$\Circled{1}$ Quant. loss mask init (\textbf{proposed}) &0.1  & D+G1 & \textbf{2.92} &\textbf{2.97} & \textbf{3.02} & \textbf{3.55} & \textbf{3.08} & \textbf{0.76} & \underline{0.80} &\textbf{0.81} &\underline{0.52} &81.84 \\
~$\Circled{2}$ Random mask init &0.1   & D+G1 &\underline{2.90} &\underline{2.95} &\underline{3.01} &\underline{3.49} &\underline{3.06} &\underline{0.75} &\underline{0.80} &\textbf{0.81} &0.51 &\textbf{82.18} \\
~$\Circled{3}$ Fully masked init &1.0  & D+G1 &2.60 & 2.54 &2.82 &3.10 &2.82 &0.72 &\textbf{0.81} &\underline{0.80} &\textbf{0.53} &81.98 \\
~$\Circled{4}$ Fully masked init &1.0  & D+G5 &2.74 & 2.74 &2.93 &3.40 &3.03 &0.73 &\textbf{0.81} &\textbf{0.81} &0.50 &81.76 \\
~$\Circled{5}$ Fully masked init &1.0  & D+G10 &2.73 & 2.76 &2.93 &3.41 &3.04 &0.73 &\textbf{0.81} &\textbf{0.81} &0.49 &\underline{82.02} \\ \hline\hline
\texttt{DisContSE}  & & & & & & & & & & & &   \\ 
~$\Circled{1}$ Quant. loss mask init (\textbf{proposed}) &0.1  & D+G1 & \underline{2.92} &\underline{2.97} & \underline{3.02} & \textbf{3.55} & \textbf{3.08} & \textbf{0.76} & \textbf{0.80} &\underline{0.81} &\textbf{0.52} &\textbf{81.84} \\
~$\Circled{6}$ - semantic enh &0.1  & D+G1 &2.90 &2.94 &3.01 &3.50 &3.06 &\textbf{0.76} &\underline{0.79} &\underline{0.81} &\underline{0.51} &80.22 \\
~$\Circled{7}$ - continuous enh &0.1  & G1 &2.51 &2.46 &2.78 &3.10 &2.76 &\underline{0.71} &\textbf{0.80} &0.79 &0.50 &\underline{81.74} \\
~$\Circled{8}$ - discrete \& semantic enh &-  & D & \textbf{2.93} &\textbf{3.02} & 3.01 & 3.46 & \textbf{3.08} & \textbf{0.76} &\textbf{0.80} &\textbf{0.82} &\textbf{0.52} &81.40 \\
~$\Circled{9}$ - critic  &0.1  & D+G1 &2.89 &2.95 &\textbf{3.03} &\underline{3.54} &\underline{3.07} &\textbf{0.76} &\textbf{0.80} &\underline{0.81} &\underline{0.51} &80.96 \\
\hline
\end{tabular}
}
}
\label{tab:ablation}
\end{table*}

\vspace{-2 mm}
\section{Results and Discussion}
\vspace{-2 mm}
\subsection{Main Results}
We first investigate the optimal $T<1.0$ for the single-step reverse process during \textit{inference} of the proposed $\texttt{DisContSE}$ on $\mathcal{D}^{\textrm{val}}$, following quantization error mask initialization. As shown in Fig. \ref{fig:pesq_dnsmos_T}, the proposed $\texttt{DisContSE}$ reaches the best PESQ and DNSMOS performance for $T=0.1$, thereby we choose this value for our proposed \texttt{DisContSE} method.

Table \ref{tab:performance} shows the performance comparison among the proposed $\texttt{DisContSE}$ and seven well-known generative diffusion-based speech enhancement baselines on $\mathcal{D}^{\textrm{test}}$. Tags D and G represent discriminative and generative models, respectively, while ``G$N$" represents the number $N$ of iterations during the reverse process of the generative model. Our proposed \texttt{DisContSE} can be regarded as hybrid approach (D+G$1$), as the continuous and semantic enhancement modules are of discriminative nature, while the discrete enhancement module is part of the diffusion iterative process. As we perform single-step diffusion, the latter computes only $\hat{\mathbf{X}}^\textrm{tok}_0$, which is then input to the \texttt{DAC} decoder to estimate the enhanced speech. As an ablation, Table \ref{tab:performance} also reports the continuous enhancement module's output $\hat{\mathbf{X}}^{\textrm{tok}}$ as the input to the \texttt{DAC} decoder to estimate the enhanced speech (last row), which delivers equal or mostly poorer performance than \texttt{DisContSE}. 

Analyzing the performance of \texttt{DisContSE} and the various baselines in Table \ref{tab:performance}, we observe the strength of \texttt{CRP} \cite{lay2024single}, as it delivers top performance in ESTOI, LPS, and WAcc, all three metrics aiming to quantify intelligibility and phone fidelity. So it is an approach with only few hallucinations. \texttt{CRP} seems strong over other metrics as well (PESQ, MOS), thereby securing the overall second rank among all methods. \texttt{SB} \cite{richter2024investigating}, on the other hand, is winner in (reference-free) DNSMOS and second-ranked in (reference-free) MOS (3.73), but clearly lags behind in reference-based metrics such as PESQ (2.57) and SpkSim (0.57), obviously providing waveforms quite different to the clean original. A similar observation can be made for \texttt{Universe++} \cite{scheibler2024universal}, as it leads the (reference-free) NISQA metric and also is second-ranked in reference-free MOS (3.73), however, performing much better than \texttt{SB} in PESQ (3.09 vs.\ 2.57). \texttt{SGMSE+} \cite{richter2023speech} is the strongest on speaker similarity, but ranks relatively poorly. 

Comparing our proposed \texttt{DisContSE} in Table \ref{tab:performance} to the seven baselines, we observe that is secures five top ranked metrics, including the MOS of 3.75 from the subjective listening test, and three second ranks. \texttt{DisContSE} combines top performance in waveform-related psychoacoustic metrics (PESQ, POLQA), with top performance in reference-free UTMOS, but still is second-ranked in \mbox{ESTOI} and LPS, meaning only a low degree of hallucinations in low SNR. \textit{In total, \texttt{DisContSE} achieves the best overall rank (2.36, lower is better), with runner-up \texttt{CRP} (3.36) being exactly one rank worse.}

\vspace{-3mm}
\subsection{Ablation Study}
\vspace{-1mm}
To further verify the rationality of the inference and training strategy of the proposed \texttt{DisContSE}, Table \ref{tab:ablation} shows ablation experiments on $\mathcal{D}^{\textrm{val}}$:
$\Circled{1}$ is inferred with quantization error mask initialization (strategy (i)), $\Circled{2}$ with the random mask initialization (strategy (ii)). Both employ $T=0.1$ for a single iteration, where $\Circled{1}$ performs better, and is selected as the proposed method for the initial mask generating strategy. $\Circled{3}$ to $\Circled{5}$ are inferred with fully masked initialization ($T=1.0$) with 1, 5, and 10 iterations, respectively, all with weaker performance. $\Circled{6}$ is trained and inferred without the semantic enhancement module, $\Circled{7}$ is trained and inferred without the continuous enhancement module, and $\Circled{8}$ is trained and inferred without both the discrete enhancement and semantic enhancement modules (continuous enhancement only), and is, accordingly, only of discriminative nature. $\Circled{9}$ is trained without self-critic. $\Circled{6}$ and $\Circled{9}$ are infered with quantization error mask initialization ($T=0.1$) for 1 iteration, while $\Circled{7}$ is infered with fully masked initialization ($T=1.0$) for 1 iteration. $\Circled{8}$ directly takes the continuous enhancement output $\hat{\mathbf{X}}^{\textrm{tok}}$ as the input to the \texttt{DAC} decoder to estimate the enhanced speech, and therefore corresponds to the last row in Table \ref{tab:performance} on $\mathcal{D}^{\textrm{val}}$. All models are trained for 100 K steps. We see the validity of all \texttt{DisContSE} modules confirmed. The proposed model reaches the most balanced performance on both intrusive and non-intrusive metrics.
\vspace{-4mm}
\section{Conclusions}
\vspace{-3mm}
We propose \texttt{DisContSE}, a hybrid (discriminative/generative) diffusion-based speech enhancement model employing both discrete codec tokens and continuous embeddings. A single-step reverse process is achieved in inference by reliable tokens estimated in a continuous enhancement module and our proposed quantization error mask initialization strategy. In a multi-metric ranking with seven other methods, our model is five times the best and three times the second best in a metric, overall clearly ranked in top position. 
\vspace{-3mm}
\section{Acknowledgment}
\vspace{-2mm}
Computational resources were provided by the German AI Service Center WestAI.

\vspace{-3mm}
\section{Compliance with Ethical Standards}
\vspace{-2mm}
The study was performed according to the principles of the Declaration of Helsinki wherever applicable. Ethical approval wasn't required.

\begin{spacing}{0.85}           
\bibliographystyle{IEEEtran}
\bibliography{refs.bib}
\end{spacing}
\end{document}